\documentclass[jkps,preprint,fleqn,showpacs,showkeys]{revtex4}
\usepackage{graphicx}
\usepackage{amssymb}
\usepackage{amsmath}
\usepackage{bm}
\newcommand{\be}{\begin{eqnarray}}
\newcommand{\ee}{\end{eqnarray}}
\DeclareMathOperator{\Tr}{Tr} 
\begin{document}
\setcounter{page}{0}
\title[]{Low temperature behavior of finite-size  one-dimensional Ising model and the partition function zeros}
\author{Julian \surname{Lee}}
\email{jul@ssu.ac.kr}
\thanks{Fax: +82-2-824-4383}
\affiliation{Department of Bioinformatics and Life Science, Soongsil University, Seoul
156-743}

\date[]{Received 30 July 2014}

\begin{abstract}
In contrast to the infinite chain, the low-temperature expansion of a one-dimensional free-field Ising model has a strong dependence on boundary conditions. I derive explicit formula for the leading term of the expansion both under open and periodic boundary conditions, and show they are related to different distributions of partition function zeros on the complex temperature plane. In particular, when the periodic boundary condition is imposed, the leading coefficient of the expansion grows with size, due to the zeros approaching the origin.
\end{abstract}

\pacs{64.60.De, 64.60.an, 65.40.Ba, 04.20.Jb}

\keywords{Partition function zeros, Lattice model, Finite-size systems, Low temperature}

\maketitle

\section{INTRODUCTION}
Ising model is a simplest model for describing the magnetic system, where a spin variable that takes two possible values lives on each lattice site, which interact only with its nearest neighbors\cite{nel,par}. It is a well-known fact that the one-dimensional Ising model does not exhibit phase transition at a finite temperature. In particular, only the magnetic susceptibility diverges,   and the specific heat remains finite as $T \to 0$, giving rise to a peak with a finite height called the Schottky anomaly~\cite{par,tari}. Although the partition function as well as average values of various physical quantities of the one-dimensional Ising model can be obtained analytically using transfer matrix formalism for an arbitrary lattice size $N$, usual focus of analysis has been the thermodynamics limit, where existence or absence of phase transition can be discussed. However, finite-size one-dimensional Ising model is interesting as a model of a biopolymer, where the spin variable corresponds to the direction of each bond~\cite{nel}.

In this work, I will consider one-dimensional Ising models of finite sizes, in the absence of the magnetic field. Especially, the focus will be on the dependence of the low-temperature behavior on the boundary conditions. It has been shown for the Ising chain that the finite-size scaling of the fluctuations and the distribution of magnetization depends on boundary condition in the low temperature limit\cite{racz}. It has also been noted that the leading term in the low-temperature expansion of free-field Ising chain is different from that of the infinite chain\cite{wang,saul}, if the periodic boundary condition is imposed. Here, I will explicitly derive the explicit expression of the low-temperature expansion for a free-field finite Ising chain, which exhibit a strong dependence on boundary conditions. I will then show that these distinct behaviors are related to different distributions of the zeros of the partition function on the complex temperature plane. In particular, I show that with periodic boundary, the size scaling of the leading term is directly related to the approach of the zeros toward the origin. This is in contrast to chains with open boundary where both partition function zeros and the specific heat per bond do not have any size dependence. The results suggest that the partition function zeros contain information on the system other than phase transition.
\section{The model}
The one-dimensional Ising model with $N$ lattices sites, in the absence of the magnetic field, is described by the Hamiltonian\cite{nel}
\be
H(\{\sigma_k\}) = - J \sum_{i=1}^{N-1} \sum_{j=i+1}^N \sigma_i \sigma_j  
\ee
for open boundary condition, where each $\sigma_i$ takes the value of $+1$ or $-1$. Here, $J>0$ corresponds to ferromagnet, and $J<0$ corresponds to anti-ferromagnet. For periodic boundary condition, there is an additional coupling between the last spin and the first spin, so the Hamiltonian reads\cite{par}
\be
{\cal H}(\{\sigma_k\}) = -  J \sum_{i=1}^{N-1} \sum_{j=i+1}^N \sigma_i \sigma_j - J \sigma_N \sigma_1.
\ee
The partition function $Z \equiv \sum_{\{\sigma_k\} \}} \exp (-\beta {\cal H}(\{\sigma_k\})) $ can be expressed in terms of transfer matrix\cite{par,nel}
\be
T_{\sigma \tilde \sigma} \equiv e^{\beta J \sigma \tilde \sigma}
\ee
as 
\be
Z_{\rm open} &=& {\bf v}^\dagger {\bf T}^{N-1} {\bf v} \nonumber\\
Z_{\rm periodic} &=& \Tr {\bf T}^{N}.
\ee
where the subscripts denote the corresponding boundary conditions, and $\bf v$ is the vector whose components are all 1. The partition functions can be expressed in terms of the two eigenvalues of $T$,
\be
\lambda_\pm = e^{\beta J} \pm e^{-\beta J} = y^{1/2} \pm y^{-1/2},
\ee
where $y \equiv e^{ 2 \beta  J }$. Note that for ferromagnet ($J>0$), the positive temperature region corresponds to $1 < y < \infty$, whereas for antiferromagnet ($J<0$), it is $0 < y < 1$. 
The vector $\bf v$ is the eigenvector of ${\bf T}$ for $\lambda_+$, and the partition functions are
\be
Z_{\rm open} &=& 2 {\lambda_+}^{N-1} = 2 (y^{1/2}+y^{-1/2})^{N-1} = 2 y^{-(N-1)/2} (1+y)^{N-1} \nonumber\\
Z_{\rm periodic} &=& {\lambda_+}^{N} + {\lambda_-}^N = (y^{1/2}+y^{-1/2})^N + (y^{1/2}-y^{-1/2})^N\nonumber\\
&=& y^{-N/2} \left[(y+1)^N + (y-1)^N \right] \label{part}
\ee
Note that the partition function is invariant under the transformation $y \leftrightarrow y^{-1}$ for open boundary condition, but the invariance holds only for even values of $N$ for periodic boundary condition. In the original Hamiltonian, this transformation corresponds to $J \leftrightarrow -J$, which can be compensated by change of variable $\sigma_{2j} \leftrightarrow -\sigma_{2j}$. Obviously, such an operation cannot be done on a periodic lattice with odd number of sites, where odd and even positions cannot be defined consistently.   

\begin{figure}
\includegraphics[width=8.0cm]{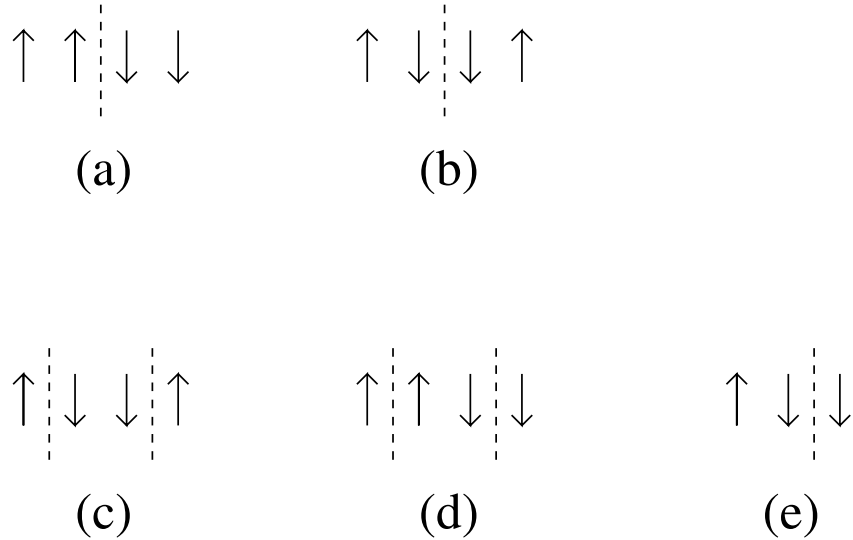}
\caption{Kinks for open chains, for (a) a ferromagnet and (b) an antiferromagnet, shown as dashed line. For periodic boundary condition the kinks appear in pairs, shown for (c) a ferromagnet and (d) an antiferromagnet. (e) A periodic antiferromagnetic chain with odd number of sites has at least one kink, so the total number of kinks is odd. 
}\label{kink}
\end{figure}

The energy is
\be
\langle E \rangle &=& -\frac{\partial}{\partial \beta} \ln Z = - 2 J y \frac{\partial}{\partial y} \ln Z= \left\{
    \begin{array}{ll}
J [ N-1 - \frac{ 2(N-1) y}{y+1} ]  &  ({\rm open}),\\
J [ N - \frac{ 2 N y \left( (y+1)^{N-1} + (y-1)^{N-1} \right) }{(y+1)^N + (y-1)^N} ]  &  ({\rm periodic})
    \end{array} \right. \label{DS}
\ee
Since the numbers of inter-spin bonds are $N-1$ and $N$ with open and periodic boundary conditions, the energies also scale with these quantities. In particular, we note that for open boundary condition, the energy is strictly proportional to $N-1$. Therefore we compare
\be
\frac{\langle E_{\rm open} \rangle}{(N-1)} &=& J [  1 - \frac{ 2  y}{y+1}]\nonumber\\
\frac{\langle E_{\rm periodic} \rangle}{N } &=& J [ 1 - \frac{ 2 y \left( (y+1)^{N-1} + (y-1)^{N-1} \right) }{(y+1)^N + (y-1)^N}] \label{Ene}
\ee
The specific heat per bond is then
\be
\frac{C_{\rm open}}{4 J^2 (N-1)} &=&   \frac{1}{k_B T^2 (N-1)}(y \frac{\partial}{\partial y})^2 \ln Z_{\rm open} = \frac{1}{k_B T^2 } \frac{y}{(y+1)^2},\nonumber\\
\frac{C_{\rm periodic}}{4 J^2 N } &=&   \frac{1}{k_B T^2 N}(y \frac{\partial}{\partial y})^2  \ln Z_{\rm periodic} \nonumber\\
&=& \frac{1}{k_B T^2 } \frac{  y \left( (y+1)^{2N-2} - (y-1)^{2N-2} + 4(N-1) y (y^2-1)^{N-2}  \right) }{[(y+1)^N + (y-1)^N]^2} \label{sh}
\ee
 Note that both energy per bond and specific heat per bond are independent of size with the open boundary condition. This is because that this model is equivalent to $N-1$ noninteracting Ising spins in a magnetic field. In fact, the elementary excitation in the free-field one-dimensional Ising model is a kink, defined as the boundary between an anti-aligned pair for ferromagnet and that between the aligned pair for the antiferromagnet, each of them costing the energy of $2 |J|$ (Fig.\ref{kink}). Since each kink can appear anywhere among $N-1$ inter-spin bonds, and the total energy is simply the sum of energy of each kink, this model can mapped into $N-1$ noninteracting Ising spins in a magnetic field, possessing two energy states per site\footnote{The difference of the free-field Ising chain from noninteracting spin chain under magnetic field is that the former has a two-fold reflection symmetry. This appears as an overall factor of 2 in the partition function and do not affect the physical quantities.}.

We also see that $(y+1)>|y-1|$ for $0 < y < \infty$,  and consequently powers of $y+1$ dominate over that of $y-1$ in the limit of $N \to \infty$ in the equations above. Therefore, we see the intensive physical quantities such as energy per bond and specific heat per bond under the periodic boundary condition converge to those under open boundary condition.
\section{Low temperature behavior}
\begin{figure}
\includegraphics[width=15.0cm]{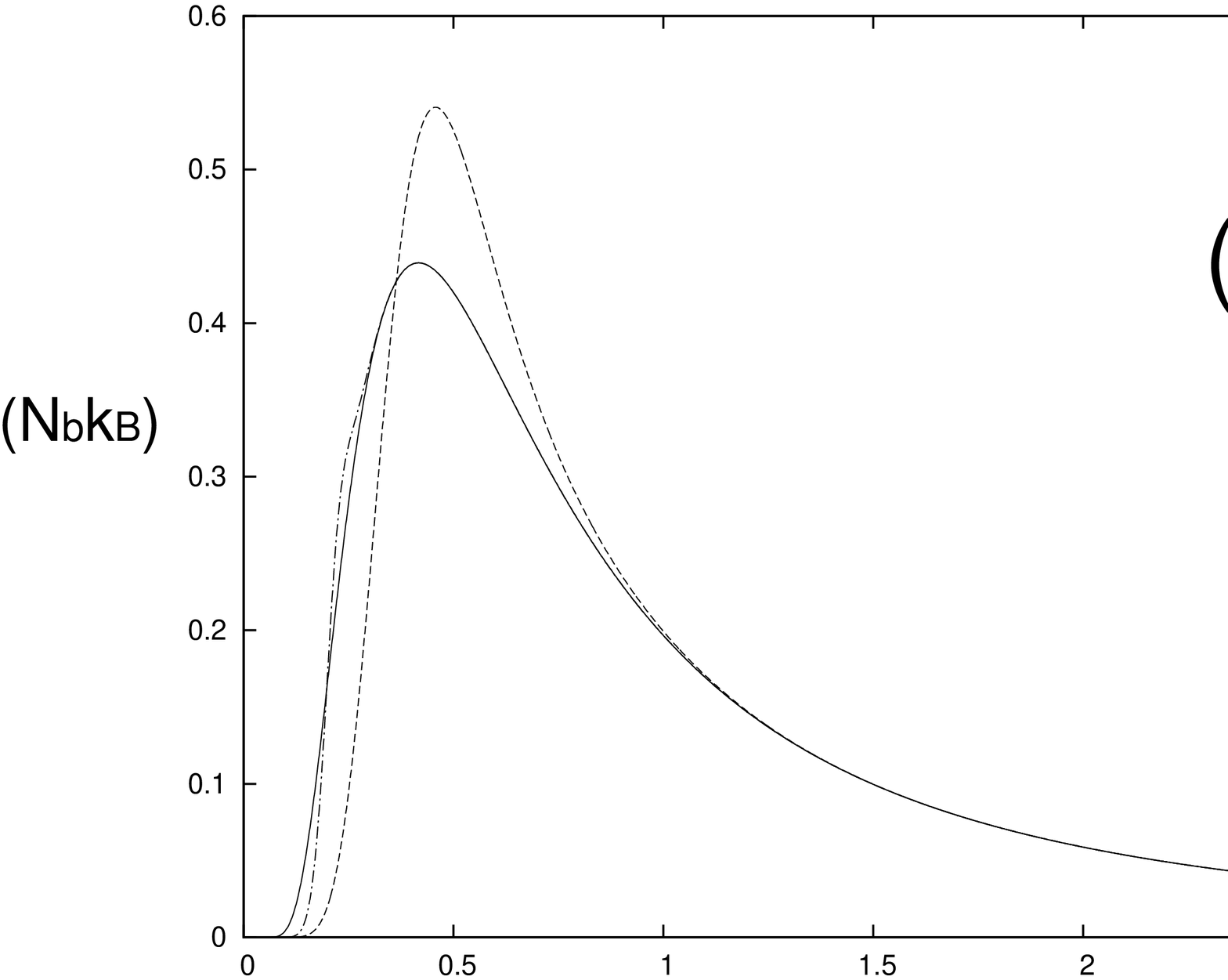}
\includegraphics[width=15.0cm]{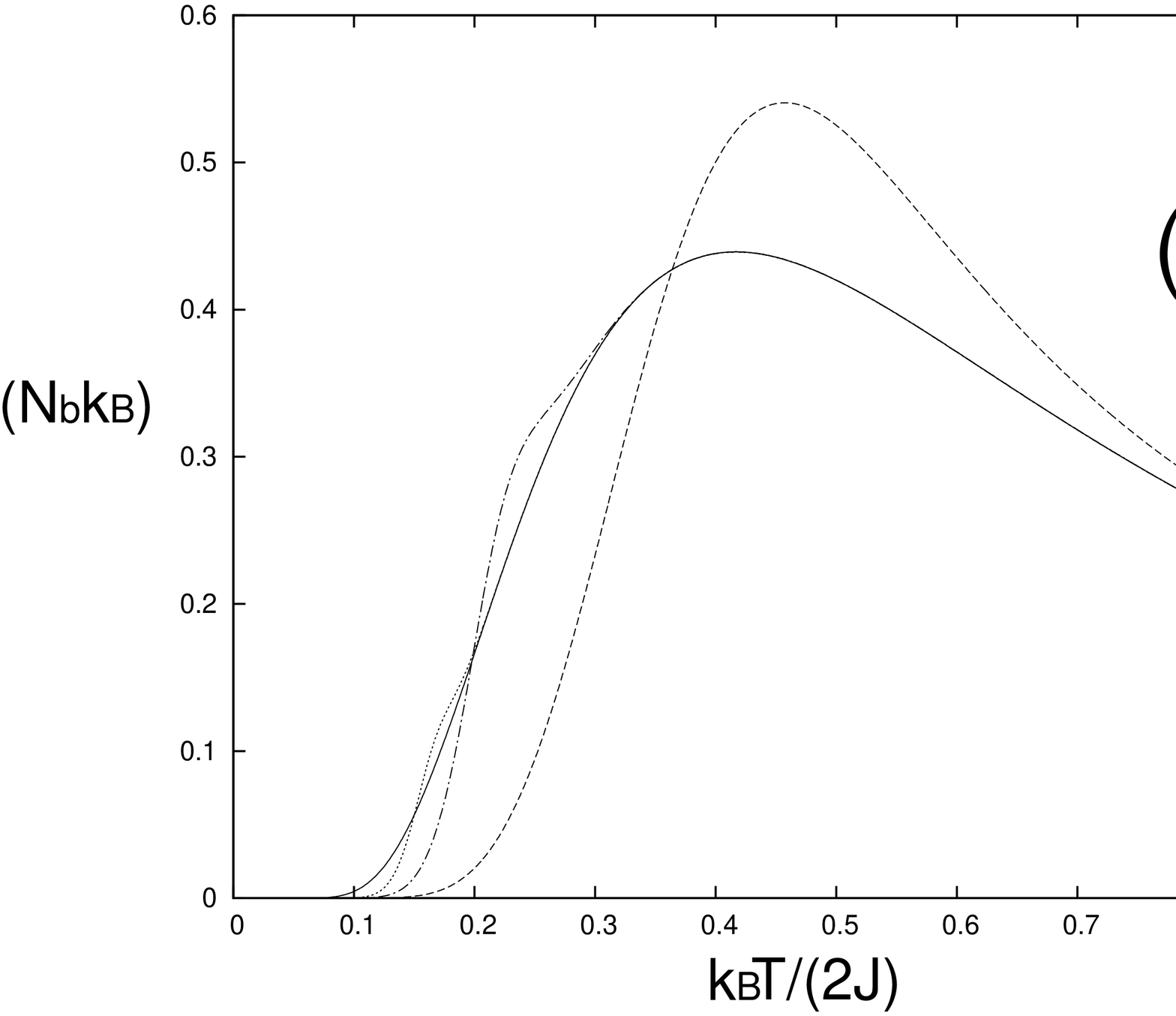}
\caption{The specific heats under periodic boundary condition for $N=10$, $N=100$ and $N=500$, compared with $N=\infty$ (open boundary).}
\label{shLs}
\end{figure}

Assuming nonnegative temperature, $T \to 0$ corresponds to  $y \to \infty$ for $J>0$ (ferromagnet) and $y \to 0$ for $J<0$ (antiferromagnet). As $y \to \infty$, the leading behavior of the specific heat is
\be
\frac{C_{\rm open} k_B T^2 }{4 J^2 (N-1)} &\sim& y^{-1} \nonumber\\
\frac{C_{\rm periodic} k_B T^2 }{4 J^2 N } &\sim&  2 (N-1) y^{-2}  \label{yinf}
\ee
On the other hand, as $y \to 0$,
\be
\frac{C_{\rm open} k_B T^2}{4 J^2 (N-1)} &\sim&  y \nonumber\\
\frac{C_{\rm periodic} k_B T^2}{4 J^2 N } &\sim&  \left\{ \begin{array}{ll}   2 (N-1) y^2 & ({\rm periodic,\ N\ even}) \\
  															  \frac{2}{3} \frac{(N-1)(N-2)}{N} y^2 & ({\rm periodic,\ N\ odd})\end{array} \right.   \label{yzero}
\ee
The results can be summarized as the as the limiting behavior as  $T \to 0$:
\be
\frac{C k_B T^2}{4 J^2 N_b} &\sim&  \left\{
    \begin{array}{ll}
     \exp(-2|J|\beta)   &   ({\rm open}),\\
       2 (N-1) \exp(-4|J|\beta) & ({\rm periodic,}\ J>0\ {\rm or\ N\ even}) \\
    \frac{2}{3}\frac{(N-1)(N-2)}{N} \exp(-4|J|\beta) & ({\rm periodic,}\ J<0\ {\rm and\ N\ odd}).
 \end{array} \right. \label{sh0}
\ee
where the number bonds is $N_b = N-1$ for open boundary and $N_b=N$ for periodic boundary.  This means that as along as $N$ is finite, even when the overall shape of intensive quantities under periodic boundary condition look very similar to that under open boundary, it will look different if we magnify the low-temperature region. An example is shown in Figure \ref{shLs}, where I show the specific heat per bond for periodic boundary for $N=10,100,500$ and compare with that for open boundary, corresponding to $N=\infty$. As shown in Fig.\ref{shLs} (a), $N=100$ approximates $N=\infty$ much better than $N=10$, but the deviation from $N=\infty$ becomes severe as $T\to 0$. As can be seen from the magnified figure, Fig.\ref{shLs}(b), the specific heat under the periodic boundary condition drops more abruptly than that for the open boundary since it falls as $e^{-4 |J| \beta}/T^2$ instead of $e^{-2 |J| \beta}/T^2$. This is why its coefficients must grow with size, in order to reduce the deviation. Divergence of this coefficient with $O(N)$ in the limit of $N \to \infty$ turns $e^{-4 |J| \beta}/T^2$ into $e^{-2 |J| \beta}/T^2$~\cite{wang}. On the other hand, in the case of the spin-glass, the coefficient grows slower than this, leading to a different low-energy behavior in the thermodynamic limit\cite{saul}. The specific heat for $N=500$ is also shown in Fig.\ref{shLs} (b), which is a much better approximation to $N=\infty$ in this temperature range. Again, the deviation from $N=\infty$ reappears if we zoom into the lower temperature region. 

These distinct low-temperature behaviors can be understood from the viewpoint of general theory of low-temperature behavior of specific heat\cite{par,tari,saul,wang}. It is clear that if the energy levels are discrete, then in the limit of $T \to 0$ only the ground state and the next excited state are relevant. Let us assume that the energy gap between the ground state and the first excited state is $\Delta E$, and their degeneracies are $g(0)$ and $g(1)$, respectively. We then have in the limit of $T \to 0$,
\be
Z \sim g(0) e^{-\beta E_0} + g(1) e^{-\beta (E_0 +\Delta E)}
\ee
and 
\be
\langle E \rangle &=& -\frac{\partial}{\partial \beta} \ln Z \sim E_0 + \frac{\omega(1) \Delta E e^{-\beta \Delta E}}{1 + \omega(1) e^{-\beta \Delta E}} = E_0 + \frac{\omega(1) \Delta E }{e^{\beta \Delta E} + \omega(1) },\nonumber\\
C &=& -\frac{1}{k_B T^2 }\frac{\partial}{\partial \beta} \langle E \rangle \sim  \frac{\omega(1) (\Delta E)^2 e^{\beta \Delta E}}{k_B T^2 \left(e^{\beta \Delta E} + \omega(1)\right)^2 } \sim  \frac{\omega(1) (\Delta E)^2 e^{-\beta \Delta E}}{k_B T^2 } \label{schott}
\ee
where $\omega(1) \equiv g(1)/g(0)$. In case of the Ising chain, there are two ground states  related by reflection symmetry, where all the spins are aligned or anti-aligned depending on whether $J>0$ (ferromagnet) or $J<0$ (antiferromagnet). Under open boundary condition, the first excited states are generated from these ground states by creation of one kink, leading to $\Delta E= 2 |J|$~(Fig.\ref{kink} (a,b)). Since there are $N-1$ places between $N$ spins to create a kink, $\omega(1) = N-1$.  For periodic boundary conditions, again there are two ground states for ferromagnet ($J>0$) and for antiferromagnet ($J<0$) with even $N$. However, this time the kinks are always created by pairs in order to satisfy the boundary condition, so the first excited state is generated by the creation of a kink pair, leading to $\Delta E= 4 |J|$~(Fig.\ref{kink} (c,d)). There are $N(N-1)/2$ places to create such a pair, so $\omega(1) = N(N-1)/2$.  Antiferromagnet ($J<0$) with odd $N$ in the presence of periodic boundary condition is quite distinct from the other cases because it is impossible to construct a state where all the spins are anti-aligned. Therefore there is at least one kink present, and since additional kinks are created in pairs, the number of kinks is always odd(Fig.\ref{kink}(e)). Again the creation of a kink pair costs $\Delta E= 4 |J|$. Counting the possible positions to place a kink, we easily see that the number of the ground states is $g(0)=2 N$, and the that of the first excited states is $g(1) = 2 N (N-1) (N-2)/6$, so we get $\omega(1) = (N-1)(N-2)/6$.  Substituting these results into Eq.(\ref{schott}) and multiplying by $k_B T^2/4 J^2 N_b$, we reproduce Eq.(\ref{sh0}). 
\section{Partition Function Zeros}
The partition function zeros are a powerful tool for studying phase transition\cite{YL,Fisher,IPZ,Bo,JK,AH,YJK,B03,Wang,CL,BDL,SYK,JKPS1,JL,Ar00,BE03,A90,prd92,prd12, Bor04,Zhu06,PRC08,PRL,JKPS2}. I will consider the zeros in the complex temperature plane. For a system that undergoes phase transition in the thermodynamic limit, the zeros approach the positive real axis as the system size grows, which is usually related to the singularity of the specific heat. Defining $z \equiv e^{-\beta \epsilon}$ with some energy scale $\epsilon$ and denoting the zeros in the complex $z$-plane as $z_i$, the partition function can be written as 
\be
Z=A(z) \prod_i (z-z_i)
\ee
where $A(z)$ is a function that is nonvanishing everywhere on the complex plane. Then the specific heat is obtained as
\be
C &=& \frac{\epsilon^2}{k_B T^2} (z\frac{\partial}{\partial z})^2 \ln Z\nonumber\\
&=& \frac{\epsilon^2}{k_B T^2} (z\frac{\partial}{\partial z}) \left( \sum_i \frac{z}{z-z_i} + \frac{z A'(z)}{A(z)} \right)\nonumber\\
&=& \frac{\epsilon^2}{k_B T^2} \left( -\sum_i \frac{z z_i}{(z-z_i)^2} + z\frac{\partial}{\partial z}\left(\frac{z A'(z)}{A(z)}\right)\right). \label{sh2}
\ee 
Therefore we see that if the zeros approach the positive real axis fast enough as the system size grows, the specific heat will blow up. To see this more clearly, we note that since the coefficient of the equation $Z(z)=0$ is real, any roots with nonzero imaginary values form complex conjugate pairs. Writing them as $z_\pm = p \pm iq$, their contribution to $k_B T^2 C/\epsilon^2$ can be written as
\be
-\frac{z z_+}{(z-z_+)^2}-\frac{z z_-}{(z-z_-)^2} &=& \frac{z[-(z_+ + z_-) z^2 + 4 z_+ z_- z - z_+ z_- (z_+ + z_-) ] }{(z-z_+)^2(z-z_-)^2}\nonumber\\
&=& \frac{z(-2 p z^2 + 4 z (p^2 + q^2) - 2 p (p^2+q^2)}{(z^2 - 2 p z + p^2 + q^2)^2} \label{shpq}
\ee
which becomes increasingly sharper near $z=p$. In particular, we see that at $z=p$, is becomes $\frac{2 p^2}{q^2}$
showing that the height of the specific heat per site $C/N$ blows up if $q$ vanishes faster than $1/N^{1/2}$.

All these standard arguments are valid only for $p>0$, and breaks down if $p=0$. That is, in contrast to zeros that approach positive real axis, those approaching the origin, $z=0$ or equivalently $T=0$, do not give rise to a singularity in the specific heat. However, as we will see in the case of the free-field Ising chain, the zeros approaching the origin gives rise to the $N$-dependence of the coefficients of the low-temperature expansion, Eq(\ref{sh0}).  

Let us first consider the free-field Ising chain under open boundary condition. We are interested in $0< y < 1$ for the case of antiferromagnet, and $1<y<\infty$ for ferromagnet, but we note that since the partition function for open boundary condition in Eq.(\ref{part}) is invariant under the inversion $y \to 1/y$, we may concentrate on  $0< y < 1$ without loss of generality. From Eq.(\ref{part}) we see that the corresponding partition function has a pole at $y=0$, which plays no role in the specific heat, since $(y \frac{\partial}{\partial y})^2 \ln y^{-N/2}=0$. On the other hand there are zeros concentrated at $y=-1$. Their positions to not change, and only their multiplicity increases with $N-1$. Therefore they do not give rise to singularities of physical quantities. They only give overall factor of $N-1$ in front of energy and specific heat, since 
\be
y \frac{\partial}{\partial y} \ln (y+1)^{N-1}&=&\frac{(N-1) y}{y+1}\nonumber\\
\left(y \frac{\partial}{\partial y}\right)^2 \ln (y+1)^{N-1}&=&-\frac{(N-1) y}{(y+1)^2}
\ee
This is to be expected, since the free-field Ising chain on $N$ sites with open boundary condition is equivalent to $N-1$ noninteracting two-state spins in a magnetic field. Extensive quantities of such noninteracting particles are simply $N-1$ times those for a single particle, and consequently intensive quantities have no $N$ dependence, leaving no room for singularity.

\begin{figure}
\includegraphics[width=15.0cm]{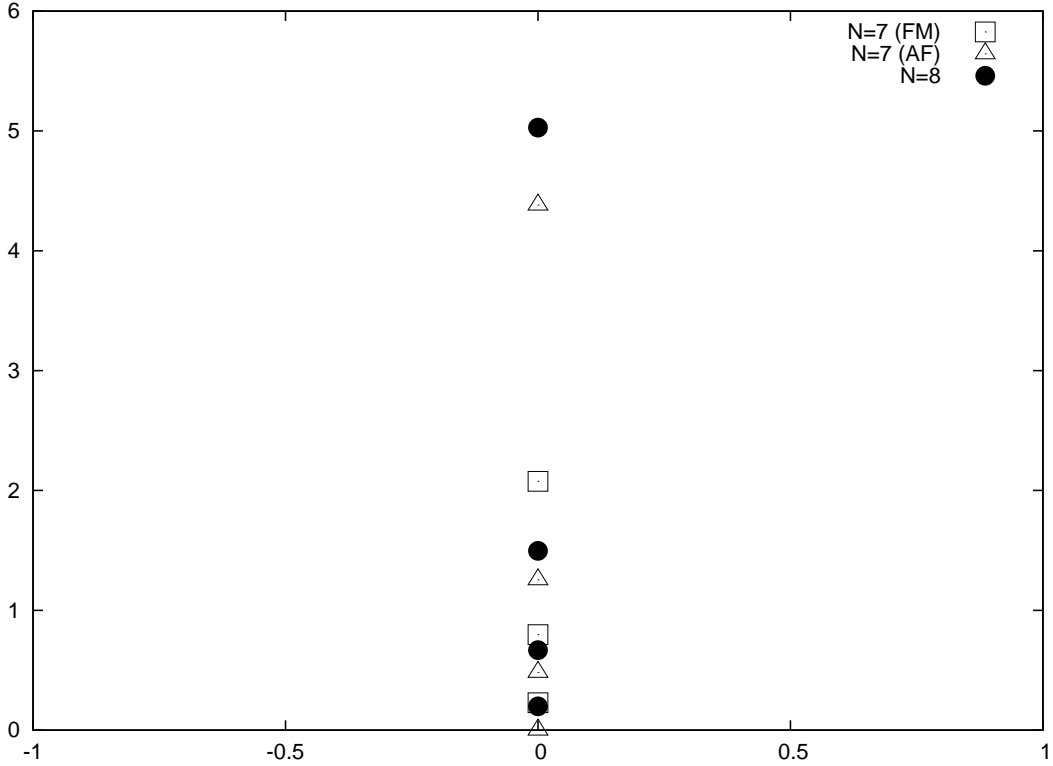}
\caption{The zeros for one-dimensional free-field Ising model under periodic boundary condition for $N=7$ and $N=8$. Only the region with nonnegative imaginary values are shown, since there is a reflection symmetry with respect to the real axis. FM and AF denote the zeros for a ferromagnet and an antiferromagnet, respectively. The pattern of zeros for FM and AF are the same for an even value of $N$. 
}\label{zero}
\end{figure}

Next we consider the free-field Ising chain under periodic boundary condition. Let us first consider the ferromagnetic case. Since $1 < y < \infty$, it is convenient to consider the complex plane of its inverse $z \equiv y^{-1}$ so that the zeros in the region $0 < z < 1$ can be investigated. In terms of $z$, partition function is written as
\be
Z_{\rm periodic} = z^{-N/2} \left[(1+z)^N + (1-z)^N \right]
\ee
Again, the pole at $z=0$ is irrelevant. Since the remaining factor is a polynomial of order $N$, $Z_{\rm periodic}$ can be rewritten as
\be
Z_{\rm periodic} = 2 z^{-N/2} \prod_{i=1}^N (z - z_i)
\ee
where $z_i$s $(i=1, \cdots, N)$ denote the $N$ zeros of the partition function. Therefore $A(z)$ dependent term in the expression Eq.(\ref{sh2}) is absent and the specific heat is proportional to the sum of the terms of the form Eq.(\ref{shpq}) evaluated at the zeros $z_i$. The zeros are obtained by solving
\be
(1+z)^N + (1-z)^N = 0.
\ee
which is equivalent to
\begin{equation}
(1+z)=\exp\left(\frac{(2j+1)i\pi}{N}\right)(1-z)
\end{equation}
with integer values of $j$,
leading to
\begin{equation}
z_j=i \tan \left(\frac{(2j+1)\pi}{2N}\right)
\end{equation}
Utilizing the relation $\tan (-\theta) = - \tan \theta$, the roots can alternatively be expressed as conjugate pairs lying on the imaginary axis:
\be
z^{\pm}_j = \pm i \tan (\frac{(2j+1)\pi}{2 N})
\ee
where $j=0 \cdots N/2-1$ for an even value of $N$, and $j=0, \cdots (N-3)/2$ for an odd value of $N$. Note that there are only $N-1$ zeros in the finite region when $N$ is odd, since the tangent function blows up for $j=(N-1)/2$. From Eq.(\ref{sh2}) with $A(z)=2z^{-N/2}$ and Eq.(\ref{shpq}) with $p=0$, we see that
\be
\frac{k_B T^2  C_{\rm periodic}}{4  J^2} &=&  4 \sum_{0 \leq j < (N-1)/2} \frac{z^2 \tan^2((2j+1)\pi/2 N)}{\left(z^2 + \tan^2((2j+1)\pi/2 N)\right)^2} \label{shz}
\ee
The leading order term as $z \to 0$ is
\be
\frac{k_B T^2  C_{\rm periodic}}{4  J^2} &\sim&    4  z^2 \sum_{0 \leq j < (N-1)/2}   \cot^{2}\frac{(2j+1)\pi}{2 N}  \label{sh0z}
\ee
To find the zeros relevant for antiferromagnets, the partition function is expressed in terms of $y$ variables:
\be
Z_{\rm periodic}
&=& y^{-N/2} \left[(y+1)^N + (y-1)^N \right] = 2 y^{-N/2} \prod_{i=1}^N (y - y_i)
\ee
where again, $y_i$s $(i=1, \cdots, N)$ denote the $N$ zeros of the partition function in the complex $y$-plane. The zeros are obtained by solving\cite{JKPS2}
\be
(y+1)^N + (y-1)^N = 0.
\ee
leading to
\begin{equation}
y_j=i \cot \frac{(2j+1)\pi}{2N}
\end{equation}
with integer values of $j$. Utilizing the relations $\cot(\theta)=\tan(\pi/2-\theta)$ and $\tan(-\theta)=-\tan \theta$, we express them as conjugate pairs of zeros lying on the imaginary axis:
\be
y^{\pm}_j = \pm i \tan \frac{(2j+1)\pi}{2 N}\quad (j=0, \cdots, N/2-1).
\ee
for even values of $N$. On the other hand, for odd value of $N$, in addition to the pair of zeros
\be
y^{\pm}_j = \pm i \tan \frac{j\pi}{N} \quad (j=1, \cdots, (N-1)/2),
\ee
there is an additional zero at the origin
\be
y_0 = 0.
\ee
The zeros for $N=7$ and $N=8$ are shown in Figure \ref{zero} as examples, where it is to be understood that the complex plane is the $z$-plane for ferromagnets and the $y$-plane for antiferromagnets. The zeros approach the origin as the size increases, which causes the coefficient of the low-energy expansion in Eq.(\ref{sh0}) grow with size, as shown next.

From Eq.(\ref{shpq}) with $p=0$, we see that the specific heat can be written as
\be
\frac{k_B T^2  C_{\rm periodic}}{4  J^2} &=&  \left\{
    \begin{array}{ll}
    4 (\sum_{j=0}^{N/2-1} \frac{y^2 \tan^2((2j+1)\pi/2 N)}{\left(y^2 + \tan^2((2j+1)\pi/2 N)\right)^2}  &   ({\rm N\ even}),\\
    4  \sum_{j=1}^{(N-1)/2} \frac{y^2 \tan^2(j\pi/N)}{\left(y^2 + \tan^2(j\pi/N)\right)^2}  &   ({\rm N\ odd})
 \end{array} \right.
 \label{shy}
\ee
Note that the zero at the origin for odd $N$ has vanishing contribution to the specific heat.  The leading order term as $y \to 0$ is
\be
\frac{k_B T^2  C_{\rm periodic}}{4  J^2} &\sim&  \left\{
    \begin{array}{ll}
    4  y^2 \sum_{j=0}^{N/2-1} \cot^{2}\frac{(2j+1)\pi}{2 N}  &   (N\ {\rm even}),\\
    4  y^2 \sum_{j=0}^{(N-3)/2} \cot^{2}\frac{(j+1)\pi}{N}  &   (N\ {\rm odd})
 \end{array} \right.
 \label{sh0y}
\ee
Summarizing Eq.(\ref{sh0z}) and (\ref{sh0y}) as behaviors near $T \to 0$, we now get
\be
\frac{k_B T^2 C_{\rm periodic}}{4 J^2} &\sim&  \left\{
    \begin{array}{ll}
       4 e^{-4|J|\beta} \left( \sum_{j=0}^{n-1} \cot^{2}\frac{(2j+1)\pi}{4 n} \right)  & (N=2n) \\
        4 e^{-4|J|\beta} \left( \sum_{j=0}^{n-1} \cot^{2}\frac{(2j+1)\pi}{(4n + 2)} \right)   & (N=2n+1\ {\rm and\ }J>0) \\
     4 e^{-4|J|\beta} \left( \sum_{j=0}^{n-1} \cot^{2}\frac{(j+1)\pi}{(2 n + 1)} \right) & (N=2n+1\ {\rm and\ }J<0).
 \end{array} \right. \label{sh0t}
\ee
where $n$ is an integer. We again confirm that the leading term is proportional to $e^{-4|J|\beta}$, with its coefficients coming from the zeros. By comparing Eq.(\ref{sh0t}) with Eq.(\ref{sh0}) for $N=2n$ and $N=2n+1$, we prove the summation formula for the series of the cotangents:
\be
\sum_{j=0}^{n-1} \cot^2\frac{(2j+1) \pi}{4n+2} &=& n (2n+1),\nonumber\\
\sum_{j=0}^{n-1} \cot^2\frac{(2j+1) \pi}{4n} &=& 3 \sum_{j=0}^{n-1} \cot^2\frac{(j+1) \pi}{2n+1} = n(2n-1). 
\ee
In particular, the growth of the coefficients of $C_{\rm periodic}$ as  $O(N^2)$ as $N \to \infty$ is seen to be directly related to the approach of the zeros to  the origin with $O(1/N)$.
\section{CONCLUSIONS}
In this work, I explicitly derived the expressions for the low-temperature expansion of one-dimensional free-field Ising models. The intensive physical quantities have no size dependence under open boundary condition, related to the fact that the positions of the partition function zeros are fixed in the complex temperature plane. Therefore, the intensive quantities are in exact agreement with those of the infinite chain, including their low-temperature behavior. On the other hand, even the intensive quantities have size-dependence under periodic boundary condition. Although these quantities approach that of the open boundary condition in the thermodynamic limit, the leading term of the low-temperature expansion is quite distinct from that of the open boundary condition.  I have shown that the size-dependence of the leading coefficient comes from the partition zeros approaching the origin on the complex temperature plane.
\begin{acknowledgments}
This work was supported by the National Research Foundation of Korea, funded by the Ministry of Education, Science, and Technology (NRF-
2012M3A9D1054705).
\end{acknowledgments}


\begin{references}

\bibitem{nel} P. Nelson, {\it Biological Physics: Energy, Information and Life} (W. H. Freeman and Co. New York 2004).
\bibitem{par} R. K. Pathria, {\it Statistical Mechanics} (Elsevier, Amsterdam 1996). 
\bibitem{tari} A. Tari, {\it The Specific Heat of Matter at Low Temperatures} (Imperial College Press, London 2003).
\bibitem{racz}
T. Antal, M. Droz, and Z. R\'acz, J. Phys. A: Math. Gen.  {\bf 37}, 1465 (2004).
\bibitem{wang}
J.-S. Wang, Phys. Rev. B {\bf 38}, 4840 (1988).
\bibitem{saul}  
L. Saul and M. Kardar Phys. Rev. E {\bf 48}, R3221 (1993).
\bibitem{YL}
C. N. Yang and T. D. Lee, Phys. Rev. {\bf 87}, 404 (1952);{\bf 87}, 410 (1952). 
\bibitem{Fisher}
M. E. Fisher, in {\it Lectures in Theoretical Physics}, Vol. 7c, edited by W. E. Brittin (University of Colorado Press, Boulder, 1965), p. 1.
\bibitem{IPZ}
C. Itzykson, R. B. Pearson, and J. B. Zuber, Nucl. Phys. B {\bf 220}, 415 (1983).
\bibitem{Bo}
P. Borrmann, O. M\"ulken, and J. Harting, Phys. Rev. Lett. {\bf 84}, 3511 (2000); O. M\"ulken, P. Borrmann, J. Harting, and H. Stamerjohanns, Phys. Rev. A {\bf 64}, 013611 (2001); O. M\"ulken and P. Borrmann, Phys. Rev. C {\bf 63}, 024306 (2001).
\bibitem{JK}
W. Janke and R. Kenna, J. Stat. Phys. {\bf 102}, 1211 (2001); Comput. Phys. Commun. {\bf 147}, 443 (2002); Nucl. Phys. B {\bf 106}, 905 (2002).
\bibitem{AH}
N. A. Alves and U. H. E. Hansmann, Phys. Rev. Lett. {\bf 84}, 1836
(2000); Physica A {\bf 292}, 509 (2001); J. Chem. Phys. {\bf 117}, 2337 (2002); Y. Peng, U. H. E. Hansmann, and N. A. Alves, J. Chem. Phys. {\bf 118}, 2374 (2003).
\bibitem{YJK} W. Janke, D. A. Johnston, and R. Kenna, Nucl. Phys. B {\bf 682}, 618 (2004); R. Kenna, D. A. Johnston, and W. A. Janke, Phys. Rev. Lett. {\bf 96}, 115701 (2006); {\bf 97} 155702 (2006). 
\bibitem{B03}
 I. Bena, F. Coppex, M. Droz, and A. Lipowski, Phys. Rev. Lett. {\bf 91}, 160602 (2003).
\bibitem{Wang}
J. Wang and W. Wang, J. Chem. Phys. {\bf 118}, 2952 (2003).
\bibitem{CL}
C.-N. Chen and C.-Y. Lin, Physica A {\bf 350}, 45 (2005).
\bibitem{BDL}
I. Bena, M. Droz, and A. Lipowski, Int. J. Mod. Phys. B {\bf 19},
4269 (2005) (and references therein).
\bibitem{SYK}
Seung-Yeon Kim, Phys. Rev. Lett. {\bf 93}, 130604 (2004); Phys. Rev. E {\bf 70,}, 016110 (2004); {\bf 71}, 017102 (2005); {\bf 74}, 011119 (2006); {\bf 81}, 031120 (2010); {\bf 82} 041107 (2010).
\bibitem{JKPS1}
J. Lee,  J. Kor. Phys. Soc. {\bf 44}, 617 (2004).
\bibitem{JL}
J. H. Lee, S.-Y. Kim, and J. Lee,  J. Chem. Phys. {\bf 133}, 114106 (2010); {\bf 135}, 204102 (2011); Phys. Rev. E {\bf 86}, 011802 (2012).  

\bibitem{Ar00}
P. F. Arndt, Phys. Rev. Lett. {\bf 84}, 814 (2000).
\bibitem{BE03}
R. A. Blythe and M. R. Evans, 

\bibitem{A90}
N. A. Alves, B. A. Berg, and S. Sanielevici, Phys. Rev. Lett. {\bf 64}, 3107 (1990). 
\bibitem{prd92}
Y. Iwasaki, K. Kanaya, and T. Yoshie, T. Hoshino and T. Shirakawa, Y. Oyanagi, S. Ichii, T. Kawai, 
Phys. Rev. D {\bf 46}, 4657 (1992)
\bibitem{prd12}
A. Bazavov, B. A. Berg, D. Du, and Y. Meurice
Phys. Rev. D {\bf 85}, 056010 (2012)

\bibitem{Bor04}
N. Borghini, R. S. Bhalerao, and J.-Y. Ollitrault, J. Phys. G: Nucl. Part. Phys. {\bf 30} (2004) S1213.
\bibitem{Zhu06}
X. Zhu, M. Bleicher, and H. St\"ocker, J. Phys. G: Nucl. Part. Phys. {\bf 32} (2006) 2181.
\bibitem{PRC08}
B. I. Abelev {\it et al.}, Phys. Rev. C, {\bf 77}, 054901 (2008).
\bibitem{PRL}
Julian Lee, Phys. Rev. Lett. {\bf110}, 248101 (2013); Phys. Rev. E {\bf 88}, 022710 (2013).
\bibitem{JKPS2}
Seung-Yeon Kim,  J. Kor. Phys. Soc. {\bf 45}, 302 (2004).  
\end{references}
\end{document}